\documentclass[aps,prl,reprint,superscriptaddress,showpacs]{revtex4-1}
\usepackage{times,graphicx,jab,subfigure,color}

\newcommand{\Alfven}{Alfv\'en}
\newcommand{\Alfvenic}{Alfv\'enic}

\begin{document}

\title{Density Fluctuation Spectrum of Solar Wind Turbulence between Ion and Electron Scales}
\author{C.~H.~K.~Chen}
\email{chen@ssl.berkeley.edu}
\author{C.~S.~Salem}
\author{J.~W.~Bonnell}
\author{F.~S.~Mozer}
\affiliation{Space Sciences Laboratory, University of California, Berkeley, California 94720, USA}
\author{S.~D.~Bale}
\affiliation{Space Sciences Laboratory, University of California, Berkeley, California 94720, USA}
\affiliation{Physics Department, University of California, Berkeley, California 94720, USA}
\begin{abstract}
We present a measurement of the spectral index of density fluctuations between ion and electron scales in solar wind turbulence using the EFI instrument on the ARTEMIS spacecraft. The mean spectral index at 1 AU was found to be --2.75 $\pm$ 0.06, steeper than predictions for pure whistler or kinetic \Alfven\ wave turbulence, but consistent with previous magnetic field measurements. The steep spectra are also consistent with expectations of increased intermittency or damping of some of the turbulent energy over this range of scales. Neither the spectral index nor the flattening of the density spectra before ion scales were found to depend on the proximity to the pressure anisotropy instability thresholds, suggesting that they are features inherent to the turbulent cascade.
\end{abstract}
\maketitle

\emph{Introduction}.---The solar wind has been observed to be a turbulent plasma for many decades (see \cite{bruno05a,horbury05,petrosyan10,matthaeus11,horbury11} for recent reviews). Understanding its properties is important for determining the universal features of turbulence and how the solar wind and collisionless plasmas in general are heated  \cite{marsch06,hollweg08,matthaeus11}. This Letter examines the properties of small scale density fluctuations in the solar wind at plasma kinetic scales, where dissipation of the turbulent energy is thought to take place.

In the inertial range, i.e., scales larger than the ion kinetic scales, the one-dimesional magnetic field power spectrum $P(k)\sim k^{\alpha}$, where $k$ is the wavenumber, is observed to have a spectral index close to $\alpha=$ --5/3 (e.g., \cite{smith06a}), consistent with a turbulent cascade. It has been known for many years \cite{behannon78,beinroth81,coroniti82,denskat83} that this spectrum steepens around ion kinetic scales, although it is still not yet obvious at which ion scale the  steepening occurs \cite{markovskii08}. More recently, a further change in the spectrum has been reported around electron scales, with either a steeper power law \cite{sahraoui09} or exponential falloff \cite{alexandrova09a} suggested. Early measurements of the spectral index between ion and electron scales showed a wide range between --4 and --1 \cite{beinroth81,denskat83,leamon98a,smith06a}, although more recently values between --2.9 and --2.3 have been obtained \cite{sahraoui09,kiyani09a,alexandrova09a,chen10b}.

The steepening at ion scales was originally attributed to ion cyclotron damping \cite{coleman68,denskat83,goldstein94} but it was later suggested that the dispersive nature of fluctuations at these scales could also be the cause \cite{ghosh96,li01,stawicki01}. It was proposed that the power law between ion and electron scales could be explained by a turbulent cascade mediated by the dispersive fluctuations \cite{galtier06b,alexandrova08b,schekochihin09} similarly to the \Alfvenic\ cascade at larger scales \cite{iroshnikov63t,kraichnan65,goldreich95}.

Theoretical predictions of the spectral index of the dispersive cascade have been made \cite{biskamp96,biskamp99a,cho04,alexandrova08b,schekochihin09} based on Kolmogorov scaling arguments \cite{kolmogorov41at}. If the turbulence is strong (non-linear eddy timescales $\approx$ linear wave timescales), the magnetic field spectral index is predicted to be --7/3. Additional effects can be included to account for the steeper observed spectra, for example, shear generated cyclotron resonant waves \cite{markovskii06}, an ion entropy cascade \cite{schekochihin09}, wave-particle scattering \cite{rudakov11}, electron Landau damping \cite{howes11a}, nonlocal interactions \cite{howes11c} or increased intermittency \cite{boldyrev12b}.

Since more than one type of plasma wave can exist at these scales, the nature of the dispersive cascade is debated. It is thought that the fluctuations may share properties of high frequency whistler waves \cite{goldstein94,stawicki01,galtier06b,gary09,chang11} or low frequency kinetic \Alfven\ waves (KAWs) \cite{leamon98a,hollweg99,bale05,howes08b,howes08a,schekochihin09,sahraoui09,chandran10b,sahraoui10,howes11a,howes11c,he12,salem12}. Since both wave modes produce turbulence with the same spectral index (--7/3), other tests have been used to distinguish between them \cite{bale05,sahraoui10,smith12,he12,salem12}. Other possible contributions to the spectrum at these scales include current sheets \cite{sundkvist07,markovskii11} and kinetic instabilities \cite{bale09} and their effect remains to be fully investigated.

The spectrum of density fluctuations has been well measured in the inertial range (e.g., \cite{marsch90b,hnat05}) but since current particle counting instruments take a several seconds to generate a density moment, it is not currently possible to measure the density spectrum below ion scales with this technique. Higher frequency measurements from the ISEE propagation experiment \cite{celnikier83} show the density spectrum flattening before the ion scales then steepening at smaller scales, although the steepening was attributed to the measurement technique. Similar spectra were seen with Cluster \cite{kellogg05} using the spacecraft potential measurement as a proxy for density (as described below), although the data resolution was not sufficient to measure far beyond the ion scales. Radio scintillation measurements also suggest a steepening of the density spectrum at ion scales in the inner solar wind \cite{coles89} and interstellar medium \cite{spangler90}.

In this Letter, we present new measurements of the solar wind density spectrum at 1 AU that have a low noise level and sufficient resolution to allow the spectral index between ion and electron scales to be determined.

\emph{Measurement technique}.---In sunlight, spacecraft emit photoelectrons and typically become positively charged. This attracts a return current of electrons from the surrounding plasma, reducing the spacecraft potential relative to the plasma, $V_{\text{sc}}$, until an equilibrium is reached in which the currents to and from the spacecraft are balanced. For higher electron density, $n_{\text{e}}$, the return current is larger, resulting in smaller $V_{\text{sc}}$. Thus, $V_{\text{sc}}$ can be used as a proxy for $n_{\text{e}}$ \cite{pedersen95}, allowing density fluctuations to be measured at a higher frequency than with particle counting instruments.

$V_{\text{sc}}$ is a good proxy for density at frequencies lower than the inverse time it takes the spacecraft to charge, which is determined by $dV_{\text{sc}}/dt=I_{\text{t}}/C$, where $C$ is the spacecraft capacitance and $I_{\text{t}}$ is the total current to the spacecraft. The important contributions to $I_{\text{t}}$ are the photoelectron current and plasma return current, $I_{\text{pl}}$, giving $I_{\text{t}}\approx I_{\text{pl}}-I_{\text{pe}} e^{-V_{\text{sc}}/T_{\text{pe}}}$, where $I_{\text{pe}}$ is the photoelectron current at $V_{\text{sc}}=0$ and $T_{\text{pe}}$ is the photoelectron e-folding energy. Applying small perturbations to the equilibrium $I_{\text{pl}}$ and $V_{\text{sc}}$, it can be shown that the spacecraft relaxes exponentially to the new equilibrium in response to density changes with time constant $\approx CT_{\text{pe}}/I_{\text{pl}}$. This corresponds to a frequency $\approx$ 6 kHz in the solar wind and in this Letter, we consider fluctuations at much lower frequencies, where $V_{\text{sc}}$ can be well calibrated to $n_{\text{e}}$.

Several intervals of high frequency data from the ARTEMIS-P2 spacecraft \cite{angelopoulos10} were used, which were in the free solar wind \cite{chen11b} and for which a reliable conversion from $V_{\text{sc}}$ to $n_{\text{e}}$ could be made. $V_{\text{sc}}$ is measured by the EFI instrument \cite{bonnell08}, which consists, in part, of four conducting spheres coupled to the plasma at the end of orthogonal booms in the spacecraft spin plane. In this Letter, data from one pair of opposite probes (probes 1 and 2) was used, since the others were found to contain large spin period spikes in their time series, likely due to shadows from one of the axial booms momentarily altering the probes' photoemission. The measured potential of the two probes relative to the spacecraft was averaged to reduce offsets due to solar wind electric fields. 

The probes themselves also charge positive from photoemission and are supplied a bias current to reduce their potential, but are left to remain about 1 V higher than the surrounding plasma. This places them at a point on their current-voltage curve where their potential is far less sensitive to density fluctuations than the spacecraft potential is. In addition to this 1 V offset, a further scale factor correction of 1.15 was applied to convert the average probe potential to a measurement of $V_{\text{sc}}$. This accounts for the fact that the probes are not infinitely far from the spacecraft but measure plasma which is slightly perturbed by the spacecraft environment (see Section 2.1 of \cite{mcfadden08a} for details of these corrections).

\begin{figure}
\includegraphics[scale=0.45]{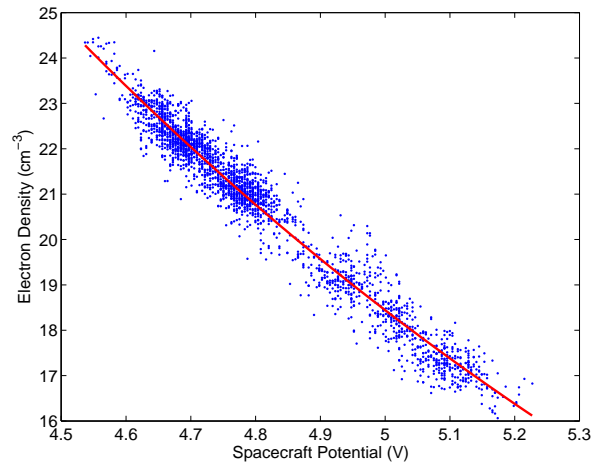}
\caption{\label{fig:cal}Electron density, $n_{\text{e}}$, as a function of the spacecraft potential relative to the solar wind, $V_{\text{sc}}$. The best fit exponential calibration curve is shown in red.}
\end{figure}

To obtain a calibration curve to convert from $V_{\text{sc}}$ to $n_{\text{e}}$, spin resolution $V_{\text{sc}}$ data was compared to $n_{\text{e}}$ data from the ESA instrument \cite{mcfadden08a}. An example of this comparison for the interval 00:04 -- 02:30 on 11th October 2010 is shown in Fig.~\ref{fig:cal}. The electron density from ESA was estimated from the measured ion density, assuming that 4\% of the ions were alphas and the rest protons (see Section 3.2.1 of \cite{mcfadden08b}). Since $n_{\text{e}}$ is expected to be proportional to the exponential of $V_{\text{sc}}$ \cite{pedersen95,scudder00}, a least squares fit of the data in Fig.~\ref{fig:cal} to the equation $n_{\text{e}}=P_1\exp\left(\left(V_{\text{sc}}-P_2\right)/P_3\right)$ was performed, where $P_1$, $P_2$ and $P_3$ are fit parameters. The $P_2$ parameter was included to allow for variations in the 1 V probe potential offset. The fit is shown as a red line in Fig.~\ref{fig:cal}.

\begin{figure}
\includegraphics[scale=0.45]{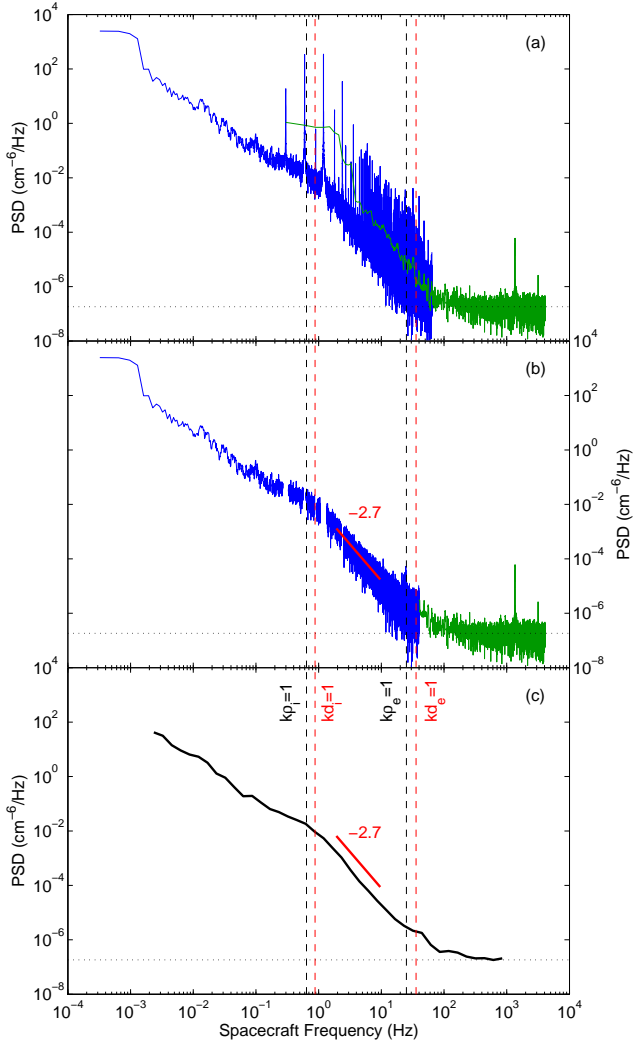}
\caption{\label{fig:spectra}Power spectra of electron density fluctuations: (a) from calibrated data (artificial spikes present), (b) with artificial spikes removed, (c) smoothed. Ion and electron gyroscales, $\rho$, and inertial lengths, $d$, are marked with vertical dashed lines. The empirical noise floor is marked with a horizontal dotted line.}
\end{figure}

\emph{Density spectrum}.---The calibration curve from Fig.~\ref{fig:cal} was applied to the 128 samples/s $V_{\text{sc}}$ data obtained during the ``particle burst'' mode interval 00:21 -- 01:14 on 11th October 2010 to obtain a density time series $n_{\text{e}}(t)$. The power spectrum of density fluctuations as a function of spacecraft-frame frequency is given by $P(f_{\text{sc}})=\int^\infty_{-\infty}R(\tau)e^{-2\pi if_{\text{sc}}\tau}d\tau$, which is the Fourier transform of the autocorrelation function $R(\tau)=\langle n_{\text{e}}(t)n_{\text{e}}(t+\tau)\rangle$, where the angular brackets denote an ensemble average. The power spectrum was estimated using the multitaper technique with time-bandwidth product $NW=4$ \cite{percival93} and is shown in Fig.~\ref{fig:spectra}a in blue. In the same figure, the spectrum of a 8192 samples/s ``wave burst'' mode interval 00:36:01 -- 00:36:05 on 11th October 2010 is shown in green.

Several features can be seen in the spectrum. Large spikes at harmonics of the spacecraft spin frequency (0.30 Hz) are present throughout the spectrum. These are caused mainly by the varying illumination of the grounded sections of the EFI booms as the spacecraft spins, altering the spacecraft photoemission and, therefore, the spacecraft potential \cite{bonnell08}. All intervals in this Letter were reduced to an integer number of spin periods to reduce spectral leakage from the spin harmonics. Since they are relatively localized in frequency, these harmonics were eliminated by removing up to 0.03 Hz either side of the spin harmonics, plus two additional sections: 1.07 Hz -- 2.28 Hz and 1.35 Hz -- 2.52 Hz. Similarly, spikes in the spectrum at harmonics of onboard clocks operating at 8 Hz and 32 Hz \cite{lecontel08} were removed. The resulting spectrum is shown in Fig.~\ref{fig:spectra}b. The low frequency portion of the high resolution (green) spectrum has also been removed since leakage from the spin harmonics here is large due to the short interval length. The spectrum was also smoothed by averaging in 45 logarithmically spaced bins from $2\times10^{-3}$ Hz to $1\times10^{3}$ Hz (Fig.~\ref{fig:spectra}c).

Under Taylor's hypothesis \cite{taylor38}, the measured frequency spectrum can be interpreted as a wavenumber spectrum since the spacecraft-frame frequency is $f_{\text{sc}}=kv_{\text{sw}}/(2\pi)$. This requires the fluctuation speeds to be less than the solar wind speed, which is well satisfied for \Alfvenic\ turbulence in the inertial range but may or may not be valid below ion scales. There is mounting evidence from phase speed \cite{bale05,salem12} and polarization \cite{he12} measurements that the fluctuations between ion and electron scales are KAW-like. Since KAWs are low frequency (compared to the ion cyclotron frequency), this suggests a wavenumber interpretation of the spectrum may be appropriate. An alternative view is that the fluctuations are not KAW-like \cite{smith12,bourouaine12}, in which case Taylor's hypothesis may break down.

Background plasma parameters for the interval were determined from the FGM \cite{auster08} and ESA \cite{mcfadden08a} instruments: solar wind speed $v_{\text{sw}}$ = 320 km/s, magnetic field strength $B$ = 5.5 nT, ion number density $n_{\text{i}}$ = 16 cm$^{-3}$, ion perpendicular temperature $T_{\perp\text{i}}$ = 9.0 eV, electron perpendicular temperature $T_{\perp\text{e}}$ = 11 eV, ion temperature anisotropy $(T_{\perp}/T_{\parallel})_{\text{i}}$ = 0.90 and electron temperature anisotropy $(T_{\perp}/T_{\parallel})_{\text{e}}$ = 1.0. The Doppler shifted kinetic scales calculated from these parameters are marked in Fig.~\ref{fig:spectra} (under Taylor's hypothesis) with vertical dashed lines.

In Fig.~\ref{fig:spectra}, at large scales (2$\times$10$^{-3}$ Hz -- 1$\times$10$^{-1}$ Hz) a power law spectrum can be seen that is consistent with previous measurements of the spectral index being around --5/3 \cite{marsch90b,hnat05,chen11b}. Just before the ion scales (0.1 Hz -- 0.7 Hz) the spectrum flattens, which has also been seen previously \cite{neugebauer75,celnikier83,kellogg05} and has been attributed to the turbulence becoming compressive \cite{hollweg99,chandran09c} or to pressure anisotropy instabilities \cite{neugebauer78}. The presence of these features suggests that the measurement technique is working well.

After steepening at ion scales, the spectrum flattens to a constant value for $f_{\text{sc}}>$ 100 Hz. This is roughly consistent with the expected instrumental noise level and has been marked as a dotted line in Fig.~\ref{fig:spectra}. The spike near 1 kHz is of unknown origin but is not important for this analysis. Between ion and electron scales a power law can be seen. The spectral index over the range $3<k\rho_{\text{i}}<15$, where the signal-to-noise ratio is large, was calculated from the gradient of the best fit line in log-log space and found to be --2.7. For greater accuracy, this spectral index was calculated from the spectrum in Fig.~\ref{fig:spectra}b and not the smoothed spectrum in Fig.~\ref{fig:spectra}c. Other spectral estimator techniques were used (e.g., windowed and wavelet transforms) with similar results. Since the spectrum reaches the noise floor around electron scales it is not yet possible to determine whether it steepens, flattens or remains the same here.

\begin{figure}
\includegraphics[scale=0.45]{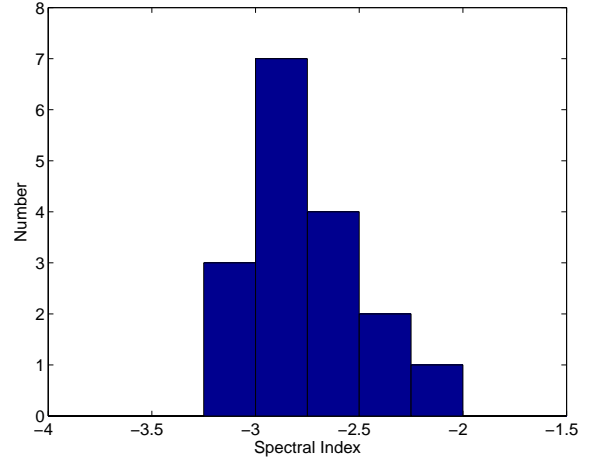}
\caption{\label{fig:hist}Histogram of electron density spectral index values between ion and electron scales ($3<k\rho_{\text{i}}<15$).}
\end{figure}

\emph{Spectral index variability}.---The same procedure was applied to 16 other intervals from October 2010 to January 2011 of duration between 6 min and 21 min. All of the intervals contained slow wind with 290 km/s $<v_{\text{sw}}<$ 350 km/s. Due to various sources of variability \cite{scudder00,pedersen08}, a different calibration curve such as in Fig.~\ref{fig:cal} was generated for each interval from a few hours of data containing the interval. A histogram of all 17 spectral indices is shown in Fig.~\ref{fig:hist}. The mean spectral index is --2.75 $\pm$ 0.06, where the error is the standard error of the mean.

\begin{figure}
\includegraphics[scale=0.45]{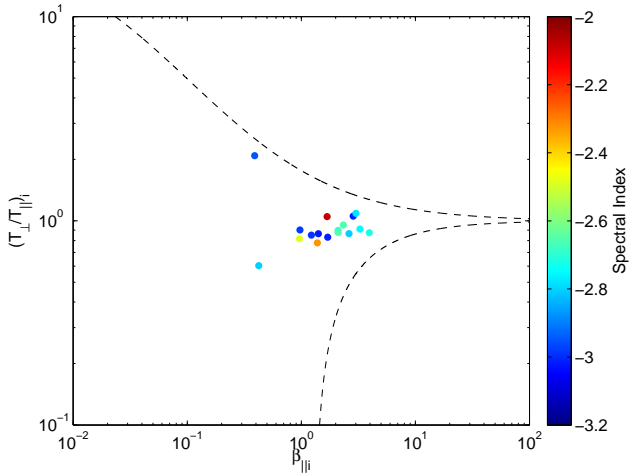}
\caption{\label{fig:instabilities}Spectral index ($3<k\rho_{\text{i}}<15$) as a function of ion temperature anisotropy and ion parallel beta. The upper dashed line is the mirror instability threshold and the lower dashed line is the oblique firehose instability threshold from \cite{hellinger06a}.}
\end{figure}

It has been noted \cite{bale09} that studies of turbulence at kinetic scales in the solar wind should consider the contribution of instability generated fluctuations to the power spectrum. For example, the power in magnetic field fluctuations at the ion gyroscale is enhanced during times when the solar wind is marginally unstable to the firehose and mirror instabilities \cite{bale09}. To examine their possible effect on the measurements in this Letter, the intervals were plotted in the instability parameter space and colored according to the spectral index (Fig.~\ref{fig:instabilities}). It can be seen that there is no consistent trend of spectral index with proximity to the thresholds, suggesting that the spectral indices measured here are not affected by these instabilities. A larger survey, with more coverage of the parameter space, however, would be required to make a more general statement.

The flattening of the density spectrum above ion scales is present in all of the intervals reported here, irrespective of their location in Fig.~\ref{fig:instabilities}. This suggests that it is inherent to the turbulent cascade, rather than being due to the pressure anisotropy instabilities. It it consistent with interpretations that the flattening is due to the compressive KAW fluctuations starting to dominate the density spectrum as the ion scales are reached \cite{hollweg99,chandran09c}.

\emph{Discussion}.---The spectral indices of density fluctuations measured here are similar to those obtained in measurements of the magnetic field at these scales (e.g., \cite{smith06a,alexandrova09a}). In particular, the mean density spectral index of --2.75 $\pm$ 0.06 is the same to within errors as the universal magnetic field spectral index of --2.8 proposed in \cite{alexandrova09a}. This is consistent with a cascade of fluctuations in which magnetic field and density are coupled and have the same spectral index, such as KAW turbulence \cite{schekochihin09}.

The measured density spectrum, however, is steeper than the prediction of --7/3 for a pure whistler or KAW cascade. This has also been seen in 3D simulations of the magnetic field spectrum of both whistler \cite{chang11} and KAW \cite{howes11a} turbulence that include kinetic effects and also a recent fluid simulation \cite{boldyrev12b}. There have been several explanations for the steep spectra, which rely on either energy being damped from the cascade or the intermittent nature of the fluctuations \cite{markovskii06,schekochihin09,rudakov11,howes11a,howes11c,boldyrev12b}. In particular, the measurement is close to the specific spectral index of --8/3 predicted in \cite{boldyrev12b}. Other possible explanations include the applicability of Taylor's hypothesis (see earlier) and anisotropy of the scaling with respect to the mean field direction \cite{chen10b}.

Recently, the density spectrum in the Earth's foreshock region has been measured at higher frequencies from 7.7 Hz to 152 Hz and the perpendicular spectrum was reported to have a spectral index of --1.6 \cite{malaspina10a}. These results cannot be directly compared to the results of this Letter or to the dispersive cascade predictions, which are for solar wind turbulence above electron scales, and these shallow large amplitude spectra remain to be explained but may be related to foreshock processes or the measurement technique.

Finally, we note that compressible turbulence in general is poorly understood, even in neutral fluids. It has recently been proposed that compressibility would cause the energy transfer rate to vary locally \cite{galtier11}, which was suggested to explain results from compressible hydrodynamic turbulence simulations \cite{kritsuk07}. We, therefore, have some way to go to fully understand the nature of compressible plasma turbulence. The measurements in this Letter place an important constraint on theoretical descriptions of such turbulence and the calibration technique will allow the possibility for more detailed analyses of solar wind density fluctuations at these scales.

\begin{acknowledgments}
This work was supported by NASA grant NNX09AE41G. We acknowledge the THEMIS/ARTEMIS team and NASA contract NAS5-02099. We thank S.~Boldyrev and J.~P.~McFadden for useful discussions.
\end{acknowledgments}

\bibliography{bibliography}

\end{document}